\documentclass[twocolumn,showpacs, superscriptaddress,aps]{revtex4}
\usepackage{graphicx}
\begin{document}
\title{Spin interference and Fano effect in electron transport through a mesoscopic ring side-coupled
with a quantum dot}
\author {Guo-Hui Ding and Bing Dong}
\affiliation{Department of Physics, Shanghai Jiao Tong University,
Shanghai, 200240, China}

\date{\today }

\begin{abstract}
We investigate the electron transport through a  mesoscopic ring
side-coupled with a quantum dot(QD) in the presence of Rashba
spin-orbit(SO) interaction.  It is shown that both the Fano
resonance and the spin interference effects  play important roles in
the electron transport properties. As the QD level is around the
Fermi energy, the total conductance shows typical Fano resonance
line shape.  By applying an electrical gate voltage to the QD, the
total transmission through the system can be strongly modulated. By
threading the mesoscopic ring with a magnetic flux, the
time-reversal symmetry of the system is broken, and a spin polarized
current can be obtained even though the incident current is
unpolarized.
 \end{abstract}
\pacs{  71.70.Ej, 73.23.-b, 72.25.-b  }
 \maketitle
\newpage
\textbf{ Introduction:}  Spin related transport in semiconductor
systems has attracted great interest in the field of
spintronics\cite{Zutic}, in which a variety of efforts are devoted
to use the electron spin instead of the electron charge for
information processing or even quantum information processing.
Several interesting spintronic devices have been proposed, the most
prominent one is the Datta-Das spin field effect
transistor(SFET)\cite{Datta}. This proposal uses the Rashba SO
coupling \cite{Bychov}to perform the controlled rotations of the
spin of electron. Although the SFET hasn't been realized in
experiment by now, it has been demonstrated in experiments that the
strength of the SO interaction is quite tunable by an external
electric field or gate voltage, which gives the SFET a promising
future\cite{Nitta}.

 The quantum coherence of electron is preserved when it is
transported through a system of mesoscopic scale. This has been
manifested explicitly in the conductance AB oscillation through the
Aharonov-Bohm(AB) interferometer\cite{Webb}, where an electrical
charge circles around the magnetic flux enclosed by the  AB ring and
acquires an AB phase. In the presence of Rashba SO coupling, aside
from the AB effect, another important Berry phase effect called the
Aharonov-Casher(AC) effect\cite{Aharonov} occurs, it is due to the
electron magnetic moment moving in an effective magnetic field
caused by the Rashba SO coupling. Recently, the AC effect has been
observed experimentally in electron transport through semiconductor
mesoscopic rings\cite{Morpurgo,Yau,Konig,Bergsten,Grbic}. The
modulation of the conductance can be explained theoretically in
terms of noninteracting electrons\cite{Konig, Frustaglia,Wang}. The
electron interaction effect on conductance in this Rashba SO
coupling ring has also been studied based on the Hubbard
model\cite{Lobos}.

   The electron transport through a AB ring with a QD embedded in
one arm of the ring is another interesting topic, and has been
studied extensively both in theory and
experiments\cite{Sun,Heary,Kobayashi}. In this case, the Coulomb
repulsive interaction in the QD has to be taken into account. When
Rashba SO interaction is presented inside the QD, It is predicted
theoretically that substantial spin-polarized current or conductance
can be induced by combined effect of the Rashba SO interaction and
the magnetic flux threading the ring\cite{Sun}.  A flux-tunable
resonant tunneling diode is proposed recently in the system with the
ring subject to Rashba SO interaction\cite{Citro}. In these AB
ring-QD systems, the important interference effect known as the Fano
resonance emerges.  For electron transport through a quantum wire
with side branches, defect levels or one dimensional rings, the Fano
effect is manifested as antiresonances in the electron transmission
coefficient\cite{Guinea,Amato, XRWang}. It is also pointed out in
Refs.\cite{Torres2006,Pastawski} that phase fluctuations due to this
kind of antiresonances can be an important road to the emergence of
decoherence.  By now, the Fano effect is known to be an ubiquitous
phenomenon observed in a large variety of systems. One important
progress in recent years is the observation of the Fano resonances
in various condensed matter systems, including an impurity atom on
metal surface\cite{Madhavan}, single-electron
transistor\cite{Gores,Johnson}, quantum dot in AB
interferometer\cite{Kobayashi}, etc.

\begin{figure}[htp]
\includegraphics[width=0.9\columnwidth,angle=270 ]{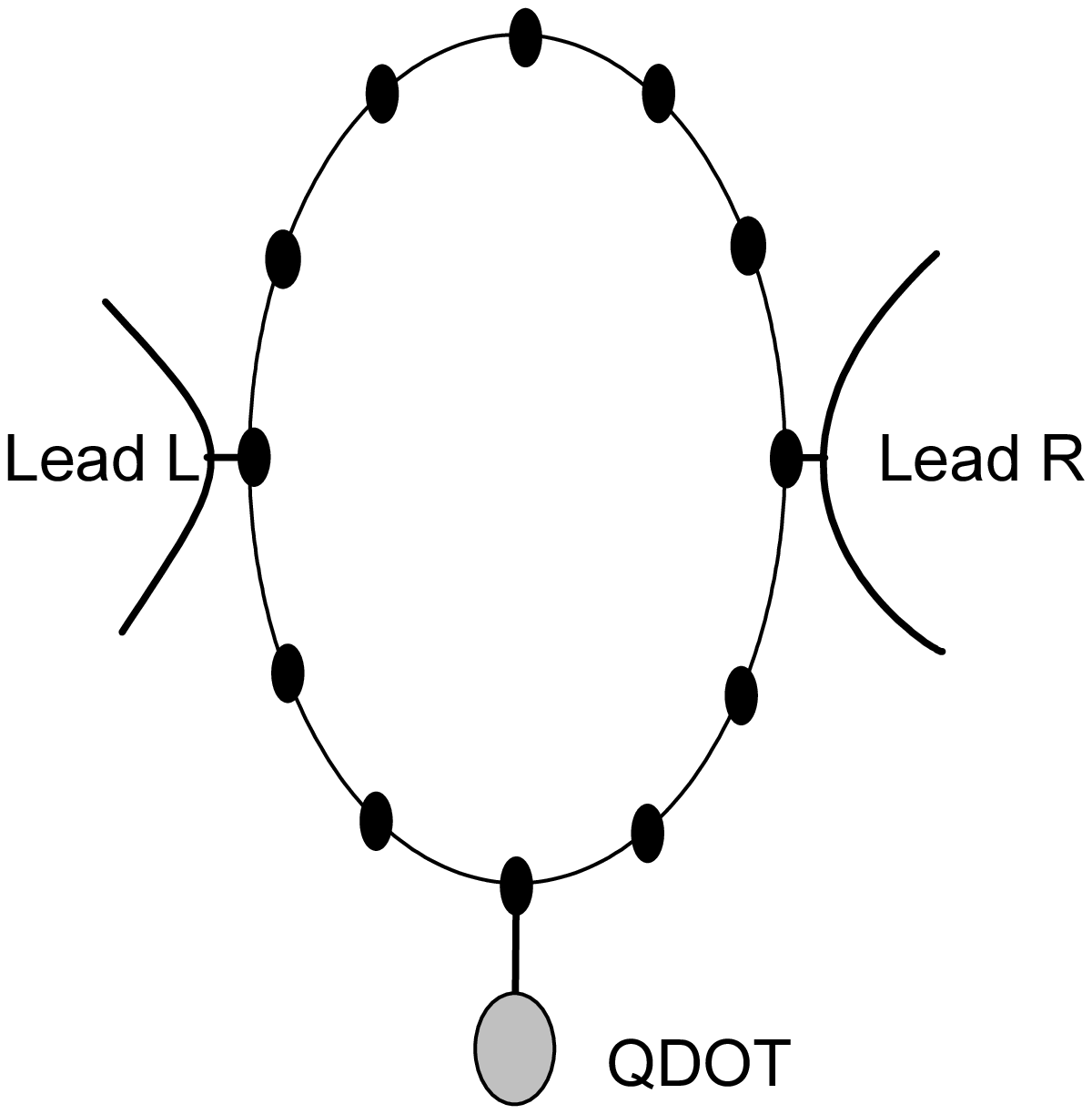}
\caption{Schematic diagram for the lattice model of two terminal AB
ring side-coupled with a quantum dot.}
\end{figure}

In this paper we  will investigate the electron transport properties
through a AB ring side-coupled with a QD, and in the presence of
Rashba SO interaction in the ring.  The geometry of this system is
plotted schematically in Fig.1.  This kind of geometry has been
realized in experiment, and  the Fano resonances in electron
transport through the ring have been observed\cite{Fuhrer}. However,
the SO coupling effects in this system haven't been well
investigated. In this paper we shall present our study on the
effects of the Rashba SO coupling on the Fano resonance and the spin
interference effects. It is shown that the interplay of the spin
interference and the Fano effect greatly influence the electron
transport properties in this system.

\textbf{Mesoscopic ring side-coupled with a quantum dot: }
 The electrons in a closed ring with
SO coupling of Rashba term can be described  by the following
Hamiltonian in the polar coordinates\cite{Meijer,Splett}
\begin{eqnarray}
&&H_{ring}=\Delta(-i{\partial\over{\partial\varphi}}+\phi)^2\nonumber\\
&& +{\alpha_R\over
2}[(\sigma_x\cos\varphi+\sigma_y\sin\varphi)(-i{\partial\over{\partial\varphi}}+\phi)+h.c.]\;,
\end{eqnarray}
where $\Delta=\hbar^2/(2m^*R^2)$, $R$ is the radius of the ring.
$\alpha_R$ will characterize the strength of Rashba SO
interaction.  $\phi=\Phi/\Phi_0$, with $\Phi$ being the external
magnetic flux enclosed by the ring, and $\Phi_0=2\pi\hbar c/e$ is
the flux quantum.

One can rewrite the above 1D Hamiltonian in terms of a lattice model
in the spatial space\cite{Souma}.
\begin{equation}
H_{ring}=-\sum_{n=1}^{N_R}
[t_\varphi^{n,n+1;\sigma,\sigma'}c^\dagger_{n,\sigma}c_{n+1,\sigma'}+H.c]\;,
\end{equation}
where $n=1, 2,\cdots, N_R$ is the lattice site index along the
azimuthal($\varphi$) direction, with $N_R$ being the total number
of the lattice sites, and we assume $N_R=4N$.
\begin{equation}
{\hat t}_\varphi^{n,n+1}=[t_0 {\hat I}_s-i
t_{so}(\sigma_x\cos\varphi_{n,n+1}+\sigma_y\sin\varphi_{n,n+1})]e^{i
2\pi\phi/N_R}\;.
\end{equation}
Here $\varphi_{n,n+1}\equiv (\varphi_n+\varphi_{n+1})/2$ with
$\varphi_n\equiv 2\pi(n-1)/N_R$, $t_0\equiv\hbar^2/2m^*a^2$ with $a$
being the lattice spacing. $t_{so}\equiv\alpha_R/2a$ is the SO
hopping parameter in this lattice model, and ${\hat I}_s$ is the
$2\times 2$ identity matrix. One can introduce a dimensionless
parameter $Q_R\equiv\alpha_R/R\Delta=(t_{so}/t_0)N_R/\pi$ to
characterize the Rashba SO interaction strength\cite{Souma}.

Now we consider the system side-coupled with a QD, which can be
described by the Anderson impurity model,
\begin{equation}
H_d=\sum_\sigma\epsilon_d d^\dagger_\sigma
d_\sigma+Un_{d\uparrow}n_{d\downarrow}\;,
\end{equation}
where $\epsilon_d$  is the energy level of QD, and $U$ is the
on-site Coulomb interaction. The tunneling between the dot level and
the ring is given by
\begin{equation}
H_{d-ring}=t_D\sum_{\sigma}(d^\dagger_\sigma
c_{N+1,\sigma}+h.c)\;.
\end{equation}
The ring is also connected to the left and right leads with the
Hamiltonian
\begin{equation}
H_{lead}=\sum_{k,\eta,\sigma}\epsilon_{k\eta}
a^\dagger_{k\eta\sigma}a^\dagger_{k\eta\sigma}\;,
\end{equation}
where  $\eta=L, R$. The tunneling Hamiltonian between the leads and
the ring is given as
\begin{eqnarray}
H_{lead-ring}&=&\sum_{k,\sigma}v_{k L}[a^\dagger_{k L
\sigma}c_{1,\sigma}+H.c] \nonumber\\
&+&\sum_{k,\sigma}v_{k R}[a^\dagger_{k R
\sigma}c_{2N+1,\sigma}+H.c]\;.
\end{eqnarray}
It should be noted that in this lattice model the electrons from the
left and right leads only have nonzero tunneling amplitudes to the
ring sites $1$ and $2N+1$, respectively.  The hybridization strength
between the lead $\eta$ and the ring can be defined as
$\Gamma^\eta=2\pi\sum_k|v_{k\eta}|^2\delta(\omega-\epsilon_{k\eta})$

Thereby, the total Hamiltonian for the system should be
\begin{equation}
H=H_{lead}+H_{ring}+H_d+H_{d-ring}+H_{lead-ring}\;.
\end{equation}

By the equation of motion method, one can obtain a set of coupled
equations for the retarded Green's functions(GFs) of the QD
$G^r_{d\sigma}$ and that of the ring $G^{r}_{i\sigma,j\sigma'}$.
However  due to the on-site Coulomb interaction, this set of
equations can't be solved exactly. We will consider the QD in the
Coulomb blockade regime. In order to treat the strong on-site
Coulomb interaction in the dot level, we use the self-consistent
decoupling procedure proposed by Hubbard\cite{Hubbard}, in which the
low energy excitations involved in the Kondo effect is neglected,
but it can properly describe the Coulomb blockade effect. Hence it
is believed to be a good approximation for temperature higher than
the Kondo temperature $T_K$. Within this approximation, the
equations of motion for the GFs have the same structure as that of
the non-interacting system, except for that now the bare Green's
function of the QD is replaced by
\begin{equation}
g^r_{d\sigma}(\omega)={{1-<n_{d\bar\sigma}>}\over{\omega-\epsilon_d}+i0^+}
+{{<n_{d\bar\sigma}>}\over{\omega-\epsilon_d-U+i0^+}}\;.
\end{equation}
Then we can obtain all retarded GFs conveniently by writing  the
Hamiltonian in the matrix form, and numerically invert the
matrix\cite{Pastawski}. As the final step, the dot occupation number
$<n_{d\bar\sigma}>$ is determined by the self-consistent equation.
\begin{equation}
<n_{d\bar\sigma}>=-\int
{d\omega\over{\pi}}n_F(\omega)ImG^{r}_{d\bar\sigma}(\omega)\;,
\end{equation}
where $n_F(\omega)$ is the Fermi-Dirac distribution function.

 For the electron transport through this system,   the spin-resolved
conductance at zero temperature will be given by the  transmission
probability at  the Fermi energy $G_{\sigma\sigma'}\equiv
T^{\sigma\sigma'}(\omega=E_F)$, with the generalized Landauer
formula\cite{Meir}
\begin{equation}
T^{\sigma\sigma'}(\omega)=Tr[{{\hat\Gamma}^\sigma}_L {\hat {G}}^r
{{\hat{\Gamma}}^{\sigma'}}_R \hat {G}^a]
\end{equation}

\textbf{Results and discussions:}
  In this section we will present the numerical results of our calculation.
We consider a quantum ring with radius $R=100
 nm$, and the effect mass $m^*=0.04 m_e$, where $m_e$ is the free
electron mass. One can obtain the energy scale $\Delta\approx 0.1
meV$. For a typical experimental accessible Rashba strength
$\alpha_R=1\times 10^{-11} eV m$, one has the parameter $\alpha_R/a
\approx \Delta$, therefore the dimensionless Rashba SO parameter
$Q_R=\alpha_R/R\Delta\approx 1$.  In our calculation,   we will take
the bandwidth parameters $t_0=1$ as the energy unit, the parameter
$t_D=0.3, U=1.0$, the total number of ring lattice sites $N_R=40$
and the Fermi energy $E_F=-0.52$.

\begin{figure}[htp]
\includegraphics[width=\columnwidth, angle=270 ]{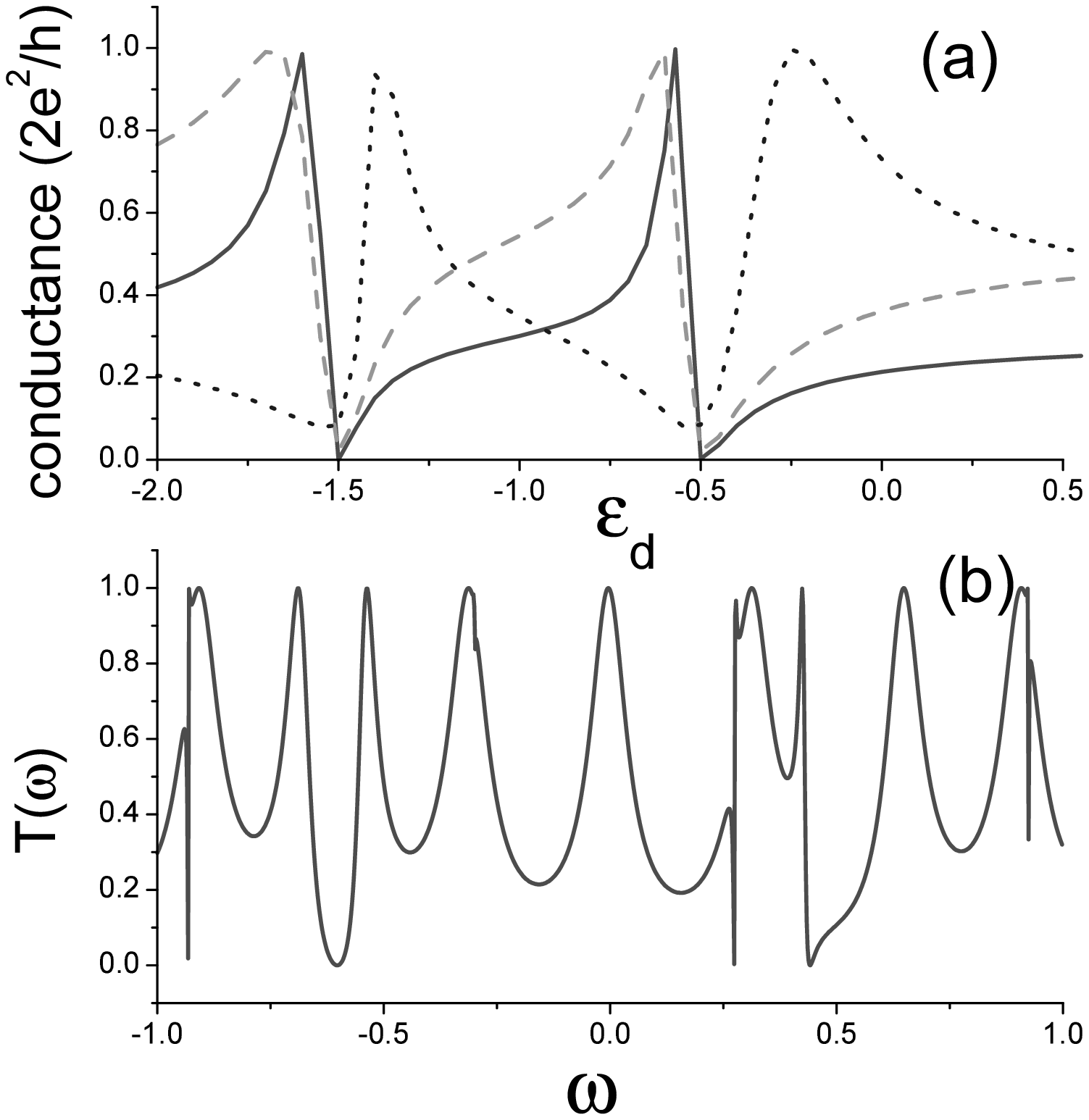}
\caption{(a) The total conductance through the 1D ring versus the QD
level $\epsilon_d$ for the Rashba SO interaction parameter
$Q_R=0.0$(solid line), $1.0$(dashed line), and $1.5$(dotted line),
respectively. (b) The transmission probability $T(\omega)$ vs the
incident electron energy for the system with quantum dot level
$\epsilon_d=-6.0$, $Q_R=0.0$.  The other model parameters are taken
as $t_0=1.0, t_D=0.5, U=1.0, E_F=-0.52$, and $\phi=0$}
\end{figure}

In Fig.2(a) the linear conductance vs. the QD level $\epsilon_d$ is
plotted at zero magnetic flux($\phi=0$), but with different SO
coupling strength($Q_R=0.0, 1.0, 1.5$).  Without the Rashba SO
coupling$(Q_R=0)$, One can see that the curve is dominated by two
peaks around $\epsilon_d-E_F\approx0$ and $\epsilon_d-E_F\approx-U$.
The peaks show typical asymmetric Fano resonance line shapes due to
the interference of electron passing through the quasi-continuous
states in ring and the discrete state in the QD.  This kind of
antiresonancs are also observed in electron transport through a
quantum wire with side-branches\cite{Guinea}, defect
levels\cite{XRWang} or one dimensional disordered rings\cite{Amato}.
It is noticed that the Rashba SO interaction can strongly affect the
Fano resonance in this system. By increasing the Rashba SO
interaction $Q_R$, the Fano line shape change drastically, for
example, the Fano resonance with $Q_R=1.5$ has opposite Fano factor
than the $Q_R=0.0 , 1.0$ cases. In order to better understand the
electron transport through the ring, we plot the transmission
probability $T(\omega)$ vs the incident electron energy in Fig.2
(b). The quasi-periodic peaks in $T(\omega)$ correspond to the
energy levels in the quantum ring. Since we assume the QD level
$\epsilon_d$ is below the Fermi energy in the calculation, the QD is
occupied by one electron, hence antiresonances are observed in the
transmission probability for electron with incident energy
$\omega\approx \epsilon_d+U$. There are also other antiresonances in
the transmission probability as shown in Fig.2(b), we will attribute
them to the fact the ring is coupled with the leads. For the system
with large bias voltage between leads, multi-levels in the ring are
involved in the electron conducting\cite{SDWang}, therefore one may
expect that the energy dependent transmission probability in the
energy range between left and right chemical potentials are
essential to the out of equilibrium transport properties of this
system. However, we shall focus our attention on the linear response
regime in the present paper.

\begin{figure}[htp]
\includegraphics[width=\columnwidth, angle=270 ]{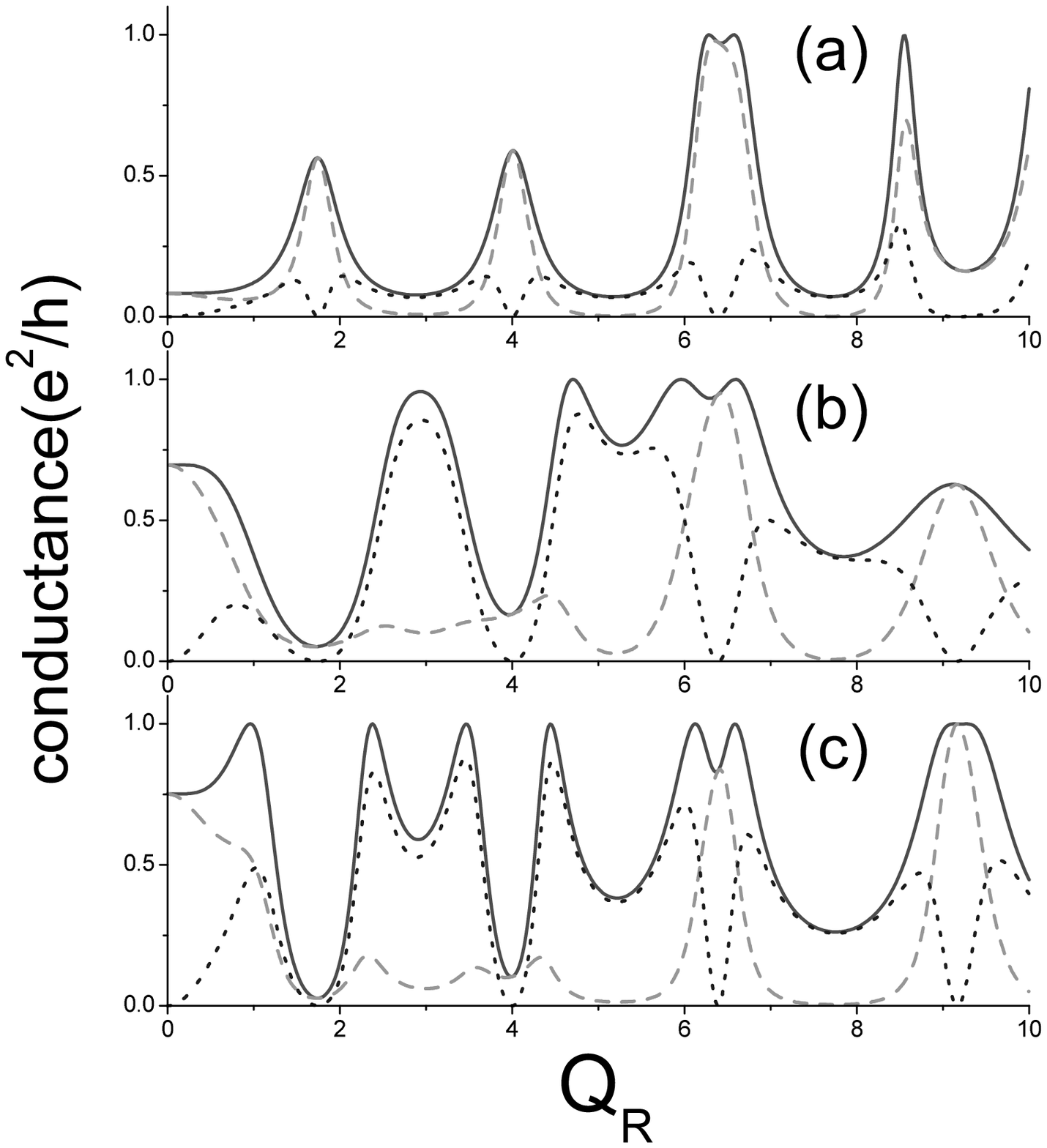}
\caption{The spin-resolved conductance versus  Rashba SO interaction
parameter for different value of dot energy level
:(a)$\epsilon_d=-0.4$; (b)$\epsilon_d=-0.55$, and (c)
$\epsilon_d=-0.6$ .  The non-spin-flip conductance
$G_{\uparrow\uparrow}$,  the spin-flip conductance
$G_{\uparrow\downarrow}$ and the spin-resolved conductance
$G_{\uparrow}=G_{\uparrow\uparrow}+G_{\uparrow\downarrow}$ are
plotted as  dashed line, dotted line and solid line, respectively.
The other parameters are the same as those of Fig.1.}
\end{figure}

Next, we consider the conductance versus Rashba SO coupling strength
for three different values of QD level $\epsilon_d=-0.4, -0.55,
-0.6$, the results are shown in Fig.3.   For the system with SO
coupling but without external magnetic flux, it still has the
time-reversal symmetry. We find the spin resolved conductances have
the symmetry $G_\uparrow=G_\downarrow$, therefore only $G_\uparrow$
is plotted. In the presence of SO interaction, the conductance shows
quasi-periodical oscillations with the value of $Q_R$ as the result
of the interference effect for electron transport through the ring.
Both the non-spin-flip $G_{\uparrow\uparrow}$ conductance and the
spin-flip conductance $G_{\uparrow\downarrow}$ can reach zero at
some particular values of $Q_R$. In this case, totally destructive
interference of this kind of spin-resolved electron transport is
achieved. It is noticed that at some value of $Q_R$, the spin
non-flip conductance $G_{\uparrow\uparrow}$ is zero, but with a
finite spin flip conductance $G_{\uparrow\downarrow}$, therefore the
ring-dot system can act as a spin flip device.  Since, the
non-spin-flip and spin-flip conductance obtain zero values  at
different position of $Q_R$,  there are always some leakage currents
in this system. It is also observed that the conductance can be
strongly modulated by the QD level, it can change from a local
maximum to a local minimum at the same value of SO coupling $Q_R$.

\begin{figure}[htp]
\includegraphics[width=\columnwidth,angle=270 ]{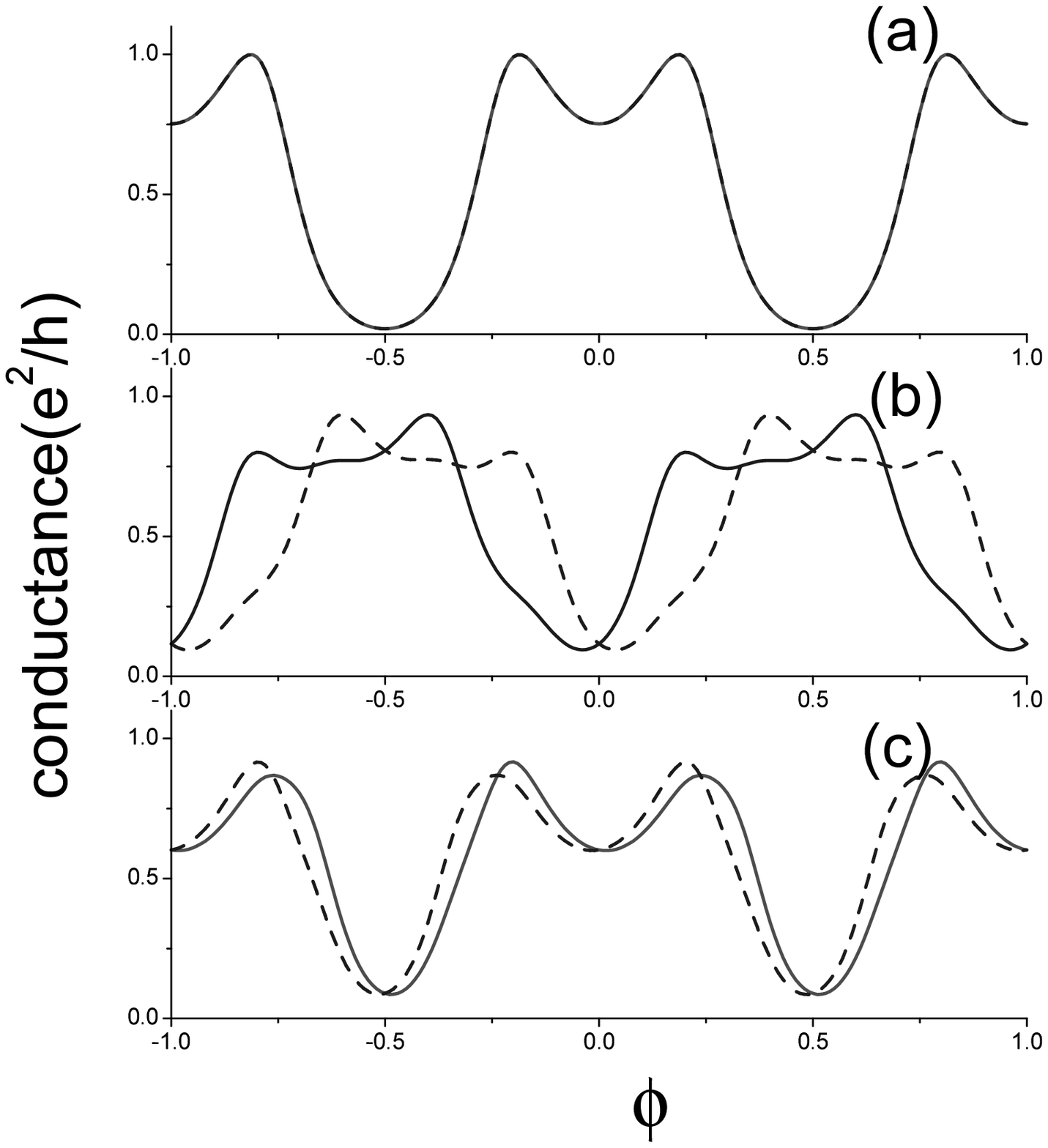}
\caption{The spin-resolved conductance $G_\uparrow$(solid line), and
$G_\downarrow$(dashed line) versus the threading magnetic flux
$\phi$ for several Rashba SO interaction parameters: (a)$Q_R=0.0$;
(b)$Q_R=1.5$; (c)$Q_R=3.0$. The dot level is fixed at
$\epsilon_d=-0.6$, and the other parameters are the same as those of
Fig.1.}
\end{figure}

In Fig.4 we give the flux dependence of the linear conductance at
different value of $Q_R$.  The conductance is a periodical function
of the enclosed magnetic flux $\phi$. Without SO coupling, the
conductance is independent of the electron spin. In the presence
both of SO interaction and  magnetic flux, the time-reversal
symmetry of this system is broken, and the result shows the
spin-resolved conductances are spin dependent ($G_\uparrow \neq
G_\downarrow$). By increasing SO coupling strength $Q_R$, the line
shapes of conductances don't change significantly, but it is easy to
observe that the conductance line shapes are shifted along the
$\phi$ axis. This can interpreted as that in the presence of SO
coupling, the electron acquires a spin Berry phase when transport
through the ring, which will act as an effective magnetic flux in
the electron conductance. Since the conductance line $G_\uparrow$
and $G_\downarrow$ shift differently, it indicates that the spin
Berry phases are different for spin up and spin down incident
electrons.

  From the above result, one can see that it is possible to obtain a spin polarized current even
the incident electron is  spin unpolarized. In the linear response
regime, we define the current spin polarization
$\eta=(G_\uparrow-G_\downarrow)/(G_\uparrow+G_\downarrow) $. For
different value of SO coupling parameter $Q_R$, the calculated
current polarization $\eta$ versus the magnetic flux $\phi$ is
plotted in Fig.5. It is observed that the spin polarization can
achieve a value of $\eta\approx 50\%$ at some particular SO coupling
strength $Q_R$ and enclosed magnetic flux $\phi$. It is interesting
to notice that at zero bias voltage, although there are no net
charge and spin currents between the left and right leads,  finite
charge and spin currents circling around the ring can exist when the
ring is threaded by a magnetic flux\cite{Ding}. It is also noted
that when the QD in this system is driven by a time dependent gate,
even at zero bias a charge or spin current might be observed due to
the quantum pumping phenomena\cite{Torres}.

\begin{figure}[htp]
\includegraphics[width=\columnwidth,angle=270 ]{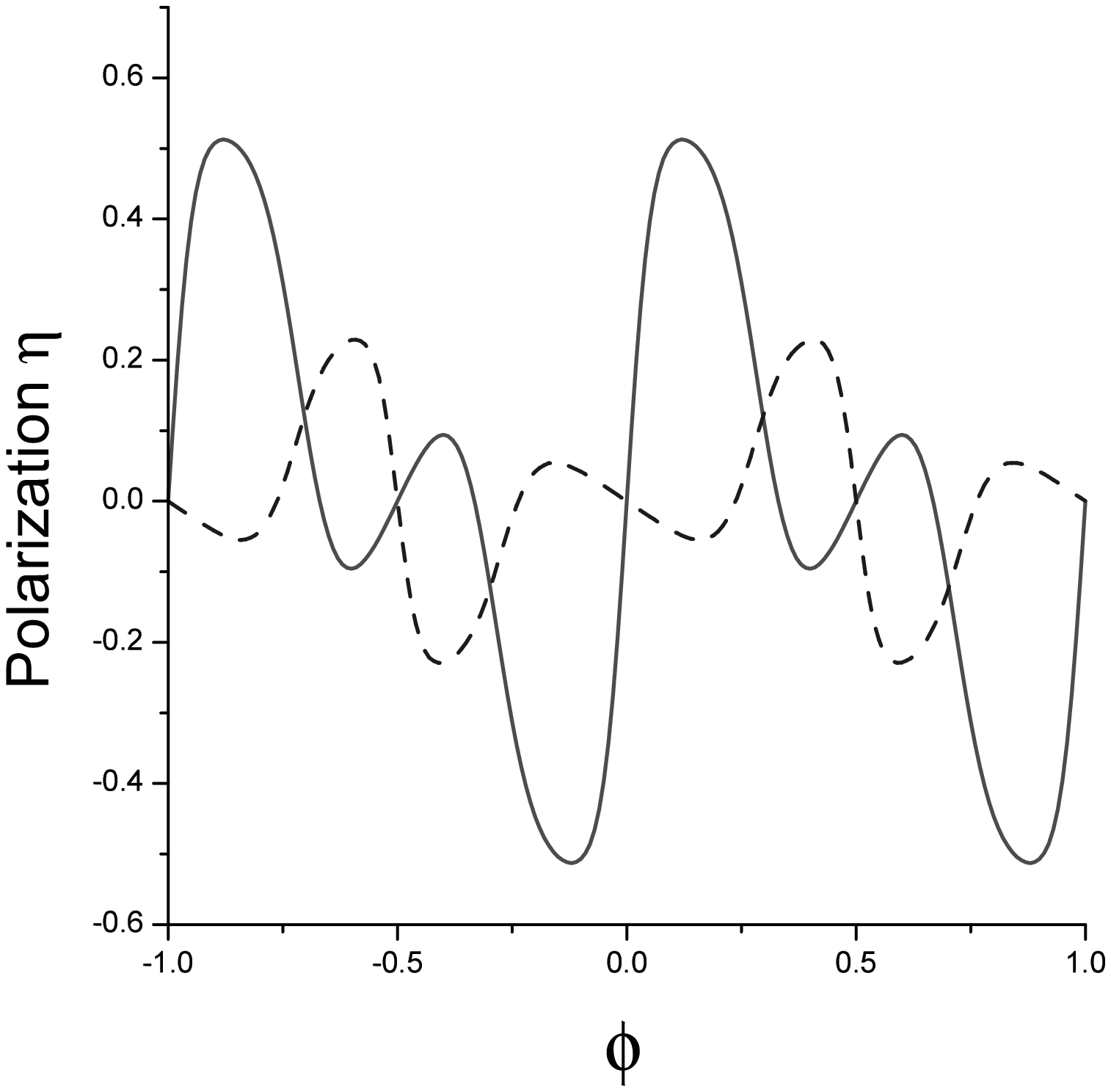}
\caption{The spin polarization $\eta$ versus the threading magnetic
flux $\phi$ with Rashba SO interaction parameters: $Q_R=1.5$(solid
line);$Q_R=3.0$(dashed line). The dot level is fixed at
$\epsilon_d=-0.6$, and the other parameters are the same as those of
Fig.1.}
\end{figure}

\textbf{Conclusions:}
      In summary, we have investigated the Rashba SO coupling effect on the
 electron transport through a AB interferometer with one
 arm side-coupled with a QD. It is shown that in the presence of a QD,  the Fano effect
 will play a  significant role in the electron transmission through the AB ring,
 therefore by tuning the QD energy level,  we can modulate the total conductance of the system.
 In the presence of Rashba SO interaction in the meososcopic ring, the electron spin can be
 reversed at some particular value of $Q_R$. With nonzero magnetic flux
 $\phi$ enclosed, the time-reversal symmetry of the system is broken,
 one can obtain a partially  spin polarized current after the electron transport through this AB ring,
 even though the incident electron is not spin polarized. The
 current spin polarization can be tuned by QD gate voltage, magnetic
 flux and the Rashba interaction strength. One may expect further
 experiment on semiconductor mesosocopic ring with Rashba SO interaction can
 reveal more interesting spin interference and Fano resonance effects.

\textbf{Acknowledgments}
 This project is supported by the National Natural Science Foundation of China, the
Shanghai Pujiang Program, and Program for New Century Excellent
Talents in University (NCET).

\end{document}